\documentclass{Interspeech2024}
\usepackage{amsfonts}
\usepackage{amsmath,graphicx, mathtools}
\usepackage{xcolor}
\usepackage{color}
\newcommand{\ci}[1]{\scriptsize{±#1}}
\definecolor{color1}{HTML}{F0F9E8}
\definecolor{color2}{HTML}{BAE4BC}
\definecolor{color3}{HTML}{7BCCC4}

\interspeechcameraready

\title{Audio Mamba: Selective State Spaces for Self-Supervised Audio Representations}

\name[affiliation={1,2}]{Sarthak}{Yadav}
\name[affiliation={1,2}]{Zheng-Hua}{Tan}

\address{
  $^1$Aalborg University, Denmark \\
  $^2$Pioneer Center for AI, Denmark 
}
\email{sarthaky@es.aau.dk, zt@es.aau.dk}


\begin{document}

\maketitle

\begin{abstract}
    Despite its widespread adoption as the prominent neural architecture,  the Transformer has spurred several independent lines of work to address its limitations. One such approach is selective state space models, which have demonstrated promising results for language modelling. However, their feasibility for learning self-supervised, general-purpose audio representations is yet to be investigated.
    This work proposes Audio Mamba, a selective state space model for learning general-purpose audio representations from randomly masked spectrogram patches through self-supervision. Empirical results on ten diverse audio recognition downstream tasks show that the proposed models, pretrained on the AudioSet dataset, consistently outperform comparable self-supervised audio spectrogram transformer (SSAST) baselines by a considerable margin and demonstrate better performance in dataset size, sequence length and model size comparisons.
\end{abstract}

\section{Introduction} 

In recent years, Transformers \cite{vaswani2017attention} and their successors have emerged as the go-to neural architecture for representation learning, traversing multiple domains and data modalities. This widespread adoption is a result of the generalization capabilities of the underlying scaled dot-product attention mechanism and transformers being inherently agnostic to input modality, as evidenced by several prominent works \cite{dosovitskiy2021an, pmlr-v162-baevski22}. The recent advancements in unsupervised and self-supervised learning have only further fueled this phenomenon, with the integration of masked predictive modeling and transformers spearheading several breakthroughs in natural language processing (NLP) \cite{bert2019}, computer vision \cite{xie2022simmim, bao2022beit, he2022masked} and audio and speech processing \cite{baevski2020wav2vec, hsu2021hubert, gong2022ssast, huang2022masked, chen2022wavlm}. However, the transformer architecture is not without it's drawbacks.
The quadratic complexity of the scaled dot-product attention operation at the heart of transformers scales poorly to very large sequences, and has spurred a lot of work on sub-quadratic approximations for attention \cite{katharopoulos2020transformers, wang2020linformer, choromanski2021rethinking} as well as token mixing approaches \cite{tolstikhin2021mlp, mai23_interspeech}. 

More recently, a new class of approaches for addressing the shortcomings of transformers has emerged, called state space models (SSMs) \cite{gu2022efficiently, nguyen2022snd, fu2023hungry, gu2023mamba}, which have demonstrated excellent long sequence modelling performance for a multitude of tasks and domains. Lying at the intersection of convolutional neural networks, recurrent neural networks and classical state space representations grounded in control theory, SSMs are sequence models governed by a set of first order differential equations. Several variants of SSMs have been proposed, ranging from (i) approaches that directly parameterize the continuous space to model time-series data \cite{zhang2023effectively}, (ii) methods that discretize SSM parameters and represent SSM computation as a convolution with a structured kernel, yielding state-of-the-art results on long sequence modelling tasks \cite{gu2022efficiently} and image and video recognition \cite{nguyen2022snd}, and most recently, (iii) context-aware adaptations of SSMs enriched by a selection mechanism that perform well on language modelling, long context speech generation \cite{gu2023mamba} as well as vision \cite{zhu2024vision} tasks. Apart from improved generalization performance on several tasks and modalities and their causal nature, SSMs and their variants have several beneficial properties, including inherent resolution invariance \cite{nguyen2022snd}, and exceptional long range modelling capabilities \cite{gu2022efficiently}.

However, while SSMs and their variants have been investigated for a wide variety of domains and tasks, a thorough assessment of their ability to learn general-purpose audio representations without supervision is pending. In this work, we propose self-supervised Audio Mamba (SSAM), an approach at the intersection of SSMs and masked predictive modelling, for learning general-purpose audio representations from randomly masked spectrogram patches. SSAMs pretrained on AudioSet \cite{gemmeke2017audio} outperform comparable self-supervised audio spectrogram transformer (SSAST) \cite{gong2022ssast} based baselines on ten downstream audio recognition tasks, consistently yielding around 30\% relative improvement in aggregate performance across several model configurations, despite having fewer number of parameters. Moreover, SSAMs adapt better to various patch sizes and input sequence lengths, while consistently performing better than SSASTs when pretrained with lower amounts of data. Code and pretrained models are available.\footnote{\url{https://github.com/SarthakYadav/audio-mamba-official}}

\section{Method}
\label{sec:method}

\subsection{Prerequisites: State space models}

\begin{figure*}
    \centering
    \includegraphics[trim={2em 7.5em 1.5em 5.5em}, clip,width=\textwidth]{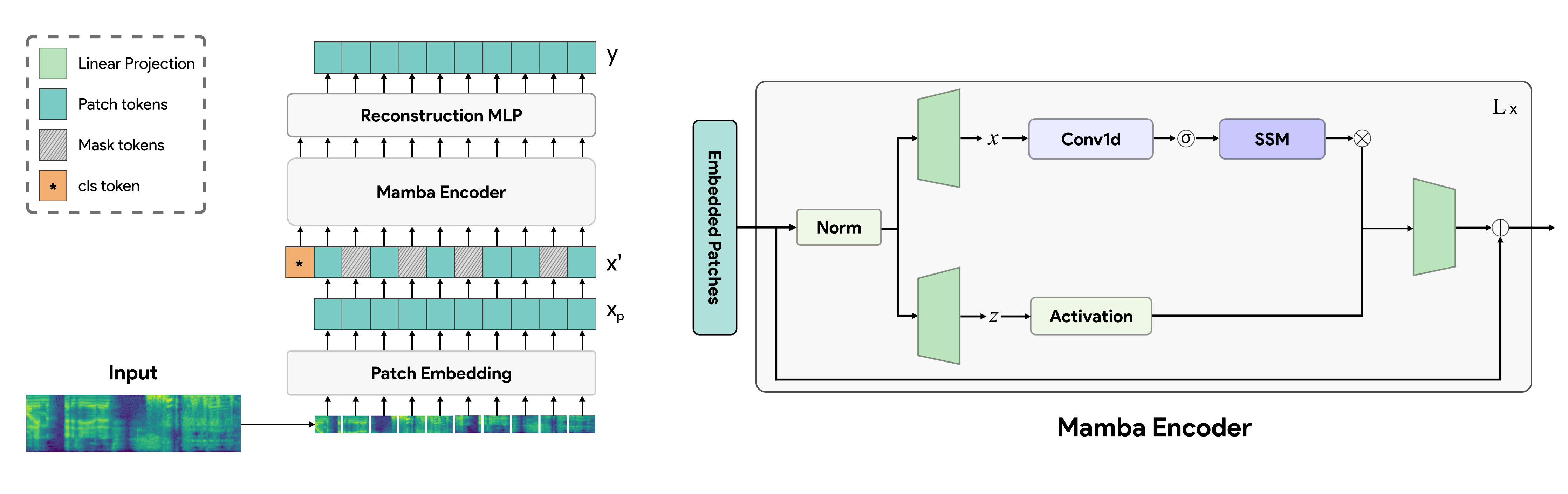}
    \caption{An overview of the proposed SSAM approach (left), and the constituent Mamba blocks (right).}
    \label{fig:ssamoverview}
\end{figure*}

Structured state space sequence models (S4) \cite{gu2022efficiently} are a recent family of linear time invariant sequence models based on a continuous system that maps input $x(t) \in \mathbb{R}$ to $y(t) \in \mathbb{R}$ through a latent state $h(t) \in \mathbb{R}^N$ using evolution parameter $\mathbf{A}$ and projection parameters $\mathbf{B}$ and $\mathbf{C}$ as follows:
\begin{align}
    h'(t) &= \mathbf{A}h(t)+\mathbf{B}x(t), \label{eq:1} \\
    y(t) &= \mathbf{C}h(t).        \label{eq:2}
\end{align}
The above equations can be discretized through a discretization rule (usually a zero-order hold), with an additional timescale parameter $\Delta$:
\begin{align}
    \overline{\mathbf{A}} &= \exp(\Delta \mathbf{A}), \\
    \overline{\mathbf{B}} &= (\Delta \mathbf{A})^{-1}(\exp(\Delta \mathbf{A}) - I)\cdot\Delta \mathbf{B}
\end{align}
Thus, S4 is effectively a discretized version of equations (\ref{eq:1}) and (\ref{eq:2}):
\begin{align}
    h_t &= \overline{\mathbf{A}}h_{t-1} + \overline{\mathbf{B}}x_t, \\
    y_t &= \mathbf{C}h_t.
\end{align}
Finally, the model SSM$(\overline{\mathbf{A}}, \overline{\mathbf{B}}, \mathbf{C})(.)$ can be computed as a global convolution between input sequence $\mathbf{x}$ and kernel $\overline{\mathbf{K}} \in \mathbb{R}^{M}$:
\begin{align}
    \label{eq:conv}
    \mathbf{\overline{K}} &= (\mathbf{C}\mathbf{\overline{B}}, \mathbf{C}\mathbf{\overline{A}}\mathbf{\overline{B}}, \dots, \mathbf{C}\mathbf{\overline{A}}^{M-1}\mathbf{\overline{B}}), \\
    \mathbf{y} &= \mathbf{x} * \mathbf{\overline{K}},
\end{align}
where ${M}$ is the length of the input $\mathbf{x}$. It's worth noting that in the above equations (5-8), the parameters for the S4 model are not conditioned on the input and are time-invariant. This is in contrast to Mamba \cite{gu2023mamba}, where parameters $\mathbf{B}, \mathbf{C} \in \mathbb{R}^{B \times L \times N}$ and $\Delta \in \mathbb{R}^{B \times L \times D}$ are instead functions of the input $\mathbf{x} \in \mathbb{R}^{B \times L \times D}$ and are context-aware, thus earning the moniker \textit{selective} structured state spaces.


\subsection{Self-Supervised Audio Mamba: SSAM}
Figure~\ref{fig:ssamoverview} (left) shows an overview of the proposed Self-Supervised Audio Mamba (SSAM) approach.
\newline
\textbf{Creating patches and random masking:} For input spectrogram $\mathbf{x} \in \mathbb{R}^{\mathtt{T} \times \mathtt{F}}$, we compute non-overlapping patches of shape $t \times f$, yielding $\mathbf{x}_p \in \mathbb{R}^{N \times (t \cdot f)}$ patches, where $N$ is the number of patches. We then flatten these patches and project them linearly to a $\mathbb{R}^{N \times d_m}$ dimensional space, followed by adding fixed sinusoidal positional embeddings for encoding positional information. We then add a representative class token to the beginning of the sequence, similar to \cite{dosovitskiy2021an, gong2022ssast}. We then proceed to randomly mask 50\% of the input patches using an unstructured masking strategy, which was demonstrated by \cite{gong2022ssast} to work better for audio tasks as compared to frame-based patching, and replace these masked patches with a learnable \textit{mask} token. Thus, input to the encoder is 
\begin{equation}
    \mathbf{x'} = [\mathtt{cls}, \mathbf{x}_p^1, \mathbf{x}_p^2, \dots, \mathbf{x}_p^{N}] + E_{\mathtt{pos}}
\end{equation}
\newline
\textbf{Encoding:} We now feed these partially masked patches to the Mamba \cite{gu2023mamba} encoder, as shown in Figure~\ref{fig:ssamoverview} (right). Mamba blocks internally expand the $d_m$ dimensional input patches by an expansion factor $E$, projecting them back to $d_m$ dimensions. The original Mamba paper \cite{gu2023mamba} uses an expansion factor $E=2$, which effectively means that every 2 Mamba blocks correspond to the same number of parameters as those of a standard Transformer block. To facilitate comparison with standard ViT encoder configurations, we instead use a ``wider'' Mamba block: with an expansion factor $E=3$ and larger internal dimensions ($d_{state}=24,d_{conv}=4$). This results in Mamba blocks that are closer to the number of parameters in a single transformer block and removes depth as a factor that might impact performance. This process yields encoded representations $\mathbf{z} = \mathtt{enc}(\mathbf{x}'), \mathbf{z} \in \mathbb{R}^{(N+1) \times d_m}$.
\newline
\textbf{Reconstruction:} After encoding, a single hidden layer MLP is used to reconstruct patches from encoded representation $\mathbf{z}$:
\begin{align}
    \mathbf{y}' &= \mathtt{Linear}_{(t \cdot f)}(\sigma(\mathtt{Linear}_{d_m}(\mathbf{z}))),
\end{align}
where $\mathtt{Linear}_d$ is a parameterized linear projection to dimensions $d$, and $\sigma$ denotes the GELU non-linear activation function \cite{hendrycks2016gaussian}. Removing the $\mathtt{cls}$ token from $\mathbf{y'}$, we get the reconstructed output $\mathbf{y} \in \mathbb{R}^{N \times (t.f)}$. For pretraining, we use mean-squared error between the original input patches $\mathbf{x}_p$ and the predicted reconstructions $\mathbf{y}$. This is in contrast to \cite{gong2022ssast}, which used a joint discriminative and reconstruction objective for pretraining. During downstream evaluation, random masking is removed and the reconstruction network is discarded, and the latent representation $\mathbf{z}$ is used. Further details can be found in Section~\ref{ssec:impdetails}.

\section{Experimental setup}
\label{sec:evalsetup}

\subsection{Datasets}

\textbf{Pretraining:} For pretraining, we use the AudioSet dataset \cite{gemmeke2017audio} (AS) which has roughly 2 million 10-second weakly labelled YouTube clips that span an ontology of 527 classes. With over 5000 hours of audio data, AudioSet is widely used for audio representation learning \cite{saeed2021contrastive, gong2022ssast, huang2022masked}.
\newline
\textbf{Downstream Evaluation:} Recently, the HEAR benchmark, which consists of 19 varied tasks from several audio domains, was proposed for thorough and systematic evaluation of audio representations. However, given the redundancy and demonstrated correlation in model performance amongst the proposed tasks \cite{turian_hear_2022}, to avoid doing excessive evaluations we utilize a subset of HEAR that consists of the following ten diverse tasks: Beijing Opera, Crema-D, ESC-50, LibriCount, Mridangam Stroke and Tonic, NSynth Pitch 5h, Speech Commands 5h, FSD50K and VoxLingua107.

\begin{table*}[t]
    \caption{Comparing SSAMs with popular self-supervised audio representations on the evaluated downstream tasks. We used pretrained models from cited papers to extract fixed feature vectors and conducted our own downstream experiments. For better 1v1 comparison, we trained directly comparable SSAST and SSAM models (highlighted with similar color levels). LS, AS, VP, LL stand for LibriSpeech, AudioSet, VoxPopuli and LibriLight datasets, respectively. *MW-MAE parameter count includes decoder parameters, which is discarded after pretraining.}
    \setlength\tabcolsep{0.9pt}
    \centering
    \begin{tabular}{lcccccccccccccc}
    \toprule
    \multicolumn{1}{l}{Model} & \multicolumn{1}{l}{Data} & \# Params & BO & CD & ESC-50 & LC & Mri-S & Mri-T & NS-5h & SC-5h & F50K & VL & $s(m)$ \\
    \midrule
    \multicolumn{3}{l}{\textbf{Naive Baselines}} & \multicolumn{5}{l}{}\\
    HEAR-Naive \cite{turian_hear_2022} & - & - & 52.6\ci{2.4} & 30.9\ci{0.8} & 5.8\ci{0.2} & 33.5\ci{1.1} & 38.0\ci{1.3} & 36.4\ci{1.9} & 18.6\ci{4.4} & 8.5\ci{0.4} & 7.1\ci{0.2} & 11.2\ci{0.5} & 5.2\ci{0.8}\\
    \midrule
    \multicolumn{3}{l}{\textbf{Supervised}} & \multicolumn{5}{l}{}\\
    PaSST-Base \cite{koutini22_interspeech} & AS & 86 M & 94.9\ci{0.5} & 61.0\ci{0.3} & 94.8\ci{0.3} & 60.1\ci{0.2} & 96.5\ci{0.1} & 87.6\ci{0.6} & 23.3\ci{0.9} & 66.6\ci{1.4} & 64.2\ci{0.1} & 25.5\ci{0.8} & 74.7\ci{0.4}\\

    \midrule
    \multicolumn{3}{l}{\textbf{SSL}} & \multicolumn{5}{l}{}\\
    W2V2-base \cite{baevski2020wav2vec} & LS & 94.4 M & 74.0\ci{1.0} & 46.4\ci{0.3} & 31.1\ci{0.4} & 51.2\ci{0.2} & 77.3\ci{0.2} & 55.1\ci{0.3} & 7.4\ci{0.8} & 90.8\ci{0.3} & 18.1\ci{0.1} & 35.5\ci{0.8} & 44.1\ci{0.2}\\
    W2V2-large \cite{baevski2020wav2vec} & VP & 315.4 M & 93.1\ci{0.7} & 66.9\ci{0.4} & 60.1\ci{0.5} & 62.4\ci{0.3} & 93.9\ci{0.1} & 77.4\ci{0.2} & 42.0\ci{1.0} & 87.6\ci{0.5} & 34.2\ci{0.1} & 53.6\ci{1.0} & 75.1\ci{0.4}\\
    WavLM-base \cite{chen2022wavlm} & LS & 94.4 M & 89.4\ci{0.7} & 56.3\ci{0.2} & 46.6\ci{0.4} & 63.2\ci{0.3} & 95.1\ci{0.1} & 83.4\ci{0.2} & 37.3\ci{0.8} & 57.2\ci{0.8} & 29.9\ci{0.1} & 22.6\ci{0.6} & 61.7\ci{0.2}\\
    WavLM-large \cite{chen2022wavlm} & Mix & 315.4 M & 96.4\ci{0.5} & 57.2\ci{0.2} & 47.9\ci{0.4} & 61.1\ci{0.3} & 96.8\ci{0.1} & 89.5\ci{0.1} & 53.7\ci{0.5} & 46.2\ci{0.8} & 29.0\ci{0.1} & 23.7\ci{0.9} & 65.1\ci{0.2}\\
    HuBERT-base \cite{hsu2021hubert} & LS & 94.4 M & 92.1\ci{0.6} & 70.8\ci{0.2} & 57.8\ci{0.6} & 56.5\ci{0.3} & 94.4\ci{0.1} & 84.9\ci{0.3} & 19.4\ci{0.7} & 93.2\ci{0.1} & 32.3\ci{0.1} & 61.8\ci{0.6} & 74.5\ci{0.2}\\
    HuBERT-large \cite{hsu2021hubert} & LL & 315.4 M & 94.1\ci{0.7} & 70.7\ci{0.1} & 60.3\ci{0.4} & 59.9\ci{0.2} & 95.3\ci{0.1} & 83.5\ci{0.3} & 19.3\ci{0.8} & 83.2\ci{0.7} & 31.5\ci{0.1} & 66.1\ci{0.9} & 74.4\ci{0.3}\\
    BEATs-iter3 \cite{chenbeats23} & AS & 90 M & 94.0\ci{0.8} & 67.3\ci{0.2} & 83.7\ci{0.3} & 68.0\ci{0.2} & 94.7\ci{0.1} & 95.8\ci{0.1} & 69.4\ci{0.8} & 85.2\ci{0.3} & 53.6\ci{0.2} & 38.5\ci{1.0} & 87.6\ci{0.3}\\
    AudioMAE \cite{huang2022masked} & AS & 86 M & 93.7\ci{0.6} & 68.2\ci{0.2} & 60.6\ci{0.4} & 42.2\ci{0.2} & 89.2\ci{0.2} & 86.6\ci{0.2} & 64.5\ci{0.8} & 28.6\ci{1.5} & 37.9\ci{0.1} & 29.7\ci{1.0} & 64.3\ci{0.3}\\
    MWMAE-Tiny \cite{yadav2024masked} & AS & 12.6 M* & 93.3\ci{1.0} & 64.4\ci{0.2} & 71.9\ci{0.5} & 65.5\ci{0.3} & 97.1\ci{0.1} & 97.6\ci{0.1} & 68.1\ci{0.4} & 77.0\ci{0.6} & 43.4\ci{0.1} & 28.6\ci{1.1} & 80.7\ci{0.3}\\
    MWMAE-Base \cite{yadav2024masked} & AS & 92.5 M* & 96.0\ci{0.5} & 73.1\ci{0.3} & 81.2\ci{0.4} & 68.8\ci{0.2} & 97.4\ci{0.1} & 97.9\ci{0.1} & 69.3\ci{0.6} & 90.9\ci{0.2} & 51.2\ci{0.2} & 44.2\ci{0.9} & 91.3\ci{0.2}\\
    \midrule
    \multicolumn{3}{l}{\textbf{SSAST Based}} & \multicolumn{5}{l}{}\\
    SSAST \cite{gong2022ssast} & AS+LS & 89 M & 93.4\ci{0.9} & 56.5\ci{0.2} & 68.4\ci{0.4} & 60.7\ci{0.3} & 96.7\ci{0.1} & 96.3\ci{0.1} & 66.8\ci{0.7} & 53.5\ci{1.3} & 38.2\ci{0.1} & 28.5\ci{0.9} & 73.1\ci{0.2}\\
    \colorbox{color1}{SSAST-Tiny} & AS & 5.4 M & 90.4\ci{0.7} & 46.9\ci{0.2} & 42.4\ci{0.6} & 42.7\ci{0.2} & 95.7\ci{0.1} & 94.3\ci{0.1} & 61.2\ci{0.5} & 50.6\ci{1.6} & 24.6\ci{0.1} & 13.8\ci{1.0} & 56.0\ci{0.2}\\
    \colorbox{color2}{SSAST-Small} & AS & 21.5 M & 93.2\ci{0.5} & 51.6\ci{0.2} & 50.1\ci{0.6} & 50.0\ci{0.3} & 96.2\ci{0.1} & 95.0\ci{0.1} & 63.8\ci{0.4} & 58.3\ci{1.2} & 31.6\ci{0.1} & 15.6\ci{0.7} & 63.4\ci{0.3}\\
    \colorbox{color3}{SSAST-Base} & AS & 85.7 M & 93.1\ci{0.7} & 56.0\ci{0.4} & 59.6\ci{0.7} & 52.9\ci{0.3} & 96.6\ci{0.1} & 96.2\ci{0.2} & 64.6\ci{0.8} & 66.1\ci{1.0} & 37.5\ci{0.1} & 19.2\ci{0.9} & 69.2\ci{0.3}\\
    \midrule
    \multicolumn{3}{l}{\textbf{Proposed}} & \multicolumn{5}{l}{}\\
    \colorbox{color1}{SSAM-Tiny} & AS & 4.8 M & 93.7\ci{0.8} & 61.8\ci{0.3} & 70.6\ci{0.2} & 59.2\ci{0.4} & 97.1\ci{0.1} & 94.9\ci{0.1} & 62.0\ci{0.7} & 74.8\ci{0.4} & 41.3\ci{0.2} & 27.8\ci{1.0} & 76.3\ci{0.2}\\
    \colorbox{color2}{SSAM-Small} & AS & 17.9 M & 94.0\ci{0.7} & 67.5\ci{0.2} & 78.7\ci{0.6} & 60.5\ci{0.3} & 97.5\ci{0.1} & 96.7\ci{0.1} & 66.3\ci{0.8} & 83.7\ci{0.3} & 48.5\ci{0.1} & 39.6\ci{0.7} & 84.4\ci{0.3}\\
    \colorbox{color3}{SSAM-Base} & AS & 69.3 M & 93.2\ci{1.1} & 70.3\ci{0.2} & 81.0\ci{0.3} & 63.5\ci{0.2} & 97.7\ci{0.1} & 96.9\ci{0.1} & 70.5\ci{0.5} & 87.9\ci{0.3} & 52.2\ci{0.1} & 50.4\ci{0.7} & 89.7\ci{0.3}\\
    \bottomrule
    \end{tabular}
    \label{tab:overall}
\end{table*}

\subsection{Implementation details}
\label{ssec:impdetails}

\textbf{Spectrogram features:} All datasets used are sampled at $16000$ Hz. Log-scaled mel spectrograms with a window size of $25$ ms, a hop size of $10$ ms and $F=80$ mel-spaced frequency bins, standardized on a per-instance basis, are used.\newline
\textbf{Pretraining:} Our default configuration consists of $l=12$ number of stacked Mamba blocks with a model feature dimension $d_m=192$, same as those of ViT-Tiny \cite{dosovitskiy2021an} encoder. We also evaluate Small ($d_m$=384, $l$=12) and Base ($d_m$=768, $l$=12) encoder configurations. All our models accept a [200, 80]-dim (time and frequency, resp.) input, corresponding to a randomly cropped 2-second audio clip during pretraining. By default, we use a patch embedding layer that computes ($4\times16$) shaped non-overlapping patches. 
All SSAM and their SSAST counterparts are trained for 100 epochs with a batch size of 1024 and a weight decay of 0.05. AdamW optimizer with a linear warmup for 10 epochs followed by a cosine learning rate decay schedule is used. No data augmentations were used.\newline
\textbf{Downstream evaluation:} After pretraining, for every input audio clip we extract fixed sized feature vectors independent of the input audio duration inline with the HEAR protocol by extracting features on 2-second audio chunks, concatenating and taking their mean over time. We then train a single hidden layer MLP classifier with 1024 neurons for each task, using the official \textit{hear-eval-kit} accompanying the HEAR benchmark. All experiments are repeated with 10 different random seeds.\newline
\textbf{Aggregated Performance Metric:} When evaluating multiple audio representations on a varied list of tasks, several previous works, such as SUPERB \cite{yang21c_interspeech}, have proposed using an aggregated performance metric that takes the difficulty levels of different tasks into account. Given the domain overlap, we use the aggregated normalized score as proposed by \cite{yadav2024masked}. Specifically, for a model $m$, overall score $s(m) \in [0.,100.]$ is given as:
\begin{align}
    s(m) = \frac{1}{|T|}\sum_{t \in T} \frac{x_t(m) - min_t}{max_t - min_t} * 100
\end{align}
where $x_t(m)$ denotes performance of the model $m$ on task $t$, and $min_t$ and $max_t$ represent the worst and the best performance across all models on the task, thus taking into account the relative performance amongst all evaluated representations.

\section{Results}
\label{sec:exps}

\subsection{Comparison with existing works}

Table~\ref{tab:overall} depicts how the proposed SSAM models fare against recent audio representations. We pretrained the highlighted configurations from scratch, and they are colour coded to represent direct comparisons: they represent methods of similar complexity and identical feature embedding size. It's worth noting that the sizes of the feature vectors extracted from other referred methods can vary, which is inline with recent frameworks for evaluating self-supervised audio representations \cite{turian_hear_2022, yang21c_interspeech}. While this might not be optimal, it is infeasible to retrain all pretrained representations to have the same embedding sizes, which also undermines the utility of training such large models. 
In the table \textit{SSAST \cite{gong2022ssast}} represents the official SSAST released model, which was pretrained with the masked prediction + reconstruction multitask objective. We can see that all the proposed SSAM models improve over their SSAST counterparts by considerable margin, offering an absolute improvement of $20$ points or over in the overall score. Moreover, both SSAM-Tiny and SSAM-Small outperform the official SSAST \cite{gong2022ssast} baseline in the overall score, despite having fewer parameters and only being trained with the reconstruction objective. SSAM-Base model, with an overall score of {89.7\ci{0.3}}, considerably outperforms the the SSAST-Base configuration as well as the official SSAST \cite{gong2022ssast} model while having fewer parameters. The proposed SSAMs outperform all evaluated methods apart from Multi-Window Masked Autoencoders (MWMAEs) \cite{yadav2024masked}, which are the current top performers for the selected evaluation protocol. However, it's worth noting that unlike SSAMs, MWMAEs are not causal. Investigating Mamba within a Masked Autoencoder (MAE) \cite{he2022masked} framework is far from trivial due to architectural nuances 
and would be the focus of future work. Overall, SSAMs outperform comparable standard transformer based approaches by a considerable margin and performs very favourably compared to popular audio representations, highlighting the effectiveness of selective state space models for learning general-purpose audio representations.

\subsection{Ablations}

\textbf{Patch Size:} We investigate how SSAM fare against standard SSASTs with changing number of input patches. To this end, we pretrain models (Tiny configuration only) with the following 3 patch size sizes: $(4,8)$, $(4,16)$, and $(8, 16)$. 
\begin{table}[h!]
    \caption{Patch size ablations with the ViT-Tiny configuration}
    \label{tab:patchablations}
    \setlength\tabcolsep{2pt}
    \centering
    \begin{tabular}{c|c|c|c}
        \toprule
        Model & Patch Size & \# Patches & $s(m)$\\
        \midrule
        SSAST-Tiny & $(8,16)$ & 125 & 47.3\ci{0.3} \\
        SSAM-Tiny &  $(8,16)$ & 125 & \textbf{58.2\ci{0.3}} \\
        \midrule
        SSAST-Tiny & $(4,16)$ & 250 & 56.0\ci{0.2} \\
        SSAM-Tiny &  $(4,16)$ & 250 & \textbf{76.3\ci{0.2}} \\
        \midrule
        SSAST-Tiny & $(4,8)$ & 500 & 54.1\ci{0.4} \\
        SSAM-Tiny &  $(4,8)$ & 500 & \textbf{76.0\ci{0.2}} \\
        \bottomrule
    \end{tabular}
\end{table}
These settings not only represent differing sequence lengths, but also different frequency and time resolution scenarios. From Table~\ref{tab:patchablations}, we can observe that the proposed SSAM approach consistently outperforms SSAST baselines for all evaluated patch sizes. Moreover, the performance gap between the two approaches grows as the number of patches increases, with the absolute difference in $s(m)$ between SSAM and SSAST increasing from $10.9$ to $20.3$ between the $(8,16)$ and $(4,16)$ settings. Overall, these experiments highlight the better ability of selective state space models to adapt to different frequency and time resolutions.
\newline
\textbf{Amount of pretraining data used:}
To simulate how both methodologies fare with the amount of data used for pretraining, we conduct ablations where we pretrain the default model with different fractions of the AudioSet corpus (Table~\ref{tab:pretrainingdata}).
\begin{table}[t]
    \caption{Overall downstream performance v/s amount of pretraining data used}
    \label{tab:pretrainingdata}
    \setlength\tabcolsep{2pt}
    \centering
    \begin{tabular}{c|c|c}
        \toprule
        Model & Pretraining data & $s(m)$\\
        \midrule
        SSAST-Tiny & 10\% of AS & 53.6\ci{0.4} \\
        SSAM-Tiny &  10\% of AS & \textbf{62.5\ci{0.2}} \\
        \midrule
        SSAST-Tiny & 25\% of AS & 52.9\ci{0.3} \\
        SSAM-Tiny &  25\% of AS & \textbf{70.5\ci{0.3}} \\
        \midrule
        SSAST-Tiny & 100\% of AS & 56.0\ci{0.2} \\
        SSAM-Tiny &  100\% of AS & \textbf{76.3\ci{0.2}} \\
        \bottomrule
    \end{tabular}
\end{table}
It is evident from Table~\ref{tab:pretrainingdata} that SSAMs scale much better with the amount of pretraining data, as the absolute difference in $s(m)$ between SSAMs and SSASTs continues to increase as more and more data is used for pretraining, highlighting the data efficacy of the proposed approach.
\newline
\textbf{Bidirectional SSM:} Given the causal, unidirectional nature of SSAM, which is unlike the attention mechanism in a transformer that learns relevance amongst all its input tokens, a natural question arises: would bidirectional computation of SSMs yield better results?
\begin{table}[h]
    \caption{Is bidirectional modelling helpful?}
    \label{tab:bidirectional}
    \centering
    \begin{tabular}{c|c|c}
    \toprule
    Model & SSM Type & $s(m)$\\
    \midrule
    SSAST-Tiny & None & 56.0\ci{0.2} \\
    SSAM-Tiny & Mamba & \textbf{76.3\ci{0.2}} \\
    SSAM-Tiny & Vim & 65.7\ci{0.3}\\
    \bottomrule
    \end{tabular}
\end{table}
\cite{zhu2024vision} attempted to address this question by proposing the \textit{Vim} block that ingests image patches with two separate Mamba subbranches operating in the forward and the backward direction, which improved performance considerably for semantic segmentation. Table~\ref{tab:bidirectional} shows ablations with bidirectional computation by replacing our underlying Mamba blocks with Vim \cite{zhu2024vision} blocks in our default configuration. We can see that while the bidirectional Vim block performs better than the standard Transformer block, it performs much worse than a unidirectional Mamba block by a considerable margin.
This makes intuitive sense, given that acoustic elements of interest in a spectrogram representation move unidirectionally in time, unlike images, where objects of interest lie in a coordinate plane. 
We thus conclude that, as opposed to image patches, bidirectional modelling with selective state spaces is not needed for audio recognition from spectrogram patches. 

\section{Conclusion}

This work presents Self-Supervised Audio Mamba (SSAM), a selective structured state space model based on Mamba for learning general-purpose audio representations from masked audio spectrograms through self-supervision. Using the AudioSet corpus for pretraining, we evaluate SSAMs against directly comparable Self-supervised Audio Spectrogram Transformer (SSAST) baselines on ten varied downstream audio recognition tasks in a multitude of settings. Following thorough empirical analysis, we conclude that SSAM models outperform baseline SSASTs by a large margin, attaining an absolute improvement of over 20 points in aggregate performance. SSAMs consistently perform better than their standard transformer-based counterparts when using less amount of pretraining data, adapt better to changing number of input sequence lengths, and scale better with models of different sizes, establishing the viability of state space models for learning general-purpose audio representations.



\bibliographystyle{IEEEtran}
\bibliography{mybib}

\end{document}